\documentclass{PoS}
\usepackage{booktabs}
\usepackage{enumitem}
\usepackage{amsmath}
\usepackage{bm}
\usepackage{tikz}
\usepackage[backend=bibtex,url,doi=false,firstinits=true,hyperref,style=numeric,sorting=none]{biblatex}
\AtEveryBibitem{\clearfield{title}}
\renewbibmacro{in:}{}
\newcommand{\bra}[1]{\mathinner{\langle{#1}|}}
\newcommand{\ket}[1]{\mathinner{|{#1}\rangle}}

\newcommand{\Ep}{E_{\bm q}}
\newcommand{\eval}[1]{\langle #1 \rangle}
\newcommand{\Eval}[1]{\big\langle #1 \big\rangle}

\title{Nucleon form factors and couplings with $N_{\mathrm f}=2+1$ Wilson fermions}

\ShortTitle{Nucleon form factors and couplings with $N_{\mathrm f}=2+1$ Wilson fermions}

\author{Dalibor~Djukanovic$^1$, \speaker{Tim~Harris}$^1$, Georg~von~Hippel$^2$, Parikshit~Junnarkar$^1$, Harvey~B.~Meyer$^{1,2}$, Hartmut~Wittig$^{1,2}$\\
        $^1$Helmholtz Institute Mainz\\
        Staudingerweg 18\\
        55128 Mainz\\
        $^2$PRISMA Cluster of Excellence and Institute for Nuclear Physics\\
        Johannes Gutenberg University of Mainz\\
        Johann-Joachim-Becher-Weg 45\\
        55128 Mainz\\
        E-mail: \email{harris@him.uni-mainz.de}
    }

   \abstract{We present updated results on the nucleon electromagnetic form factors and axial coupling calculated using CLS ensembles with $N_\mathrm{f}$=2+1 dynamical flavours of Wilson fermions.
   The measurements are performed on large, fine lattices with a pseudoscalar mass reaching down to 200 MeV.
   The truncated-solver method is employed to reduce the variance of the measurements.
   Estimation of the matrix elements is challenging due to large contamination from excited states and further investigation is necessary to bring these effects under control.
   }

\FullConference{34th annual International Symposium on Lattice Field Theory\\
         24-30 July 2016\\
         University of Southampton, UK}

\begin{document}

\section{Introduction}
\label{sec:introduction}
Knowledge of nucleon matrix elements of local currents and densities is necessary to constrain new interactions which do not exist in the Standard Model from neutron decays~\cite{Bhattacharya:2011qm}.
Furthermore, the nucleon matrix elements of the isovector vector current, $V_\mu(x)=\bar{\psi}(x)\gamma_\mu\frac{\tau^3}{2}\psi(x)$, and axial vector current, $A_\mu(x)=\bar{\psi}(x)\gamma_\mu\gamma_5\frac{\tau^3}{2}\psi(x)$, are related to experimentally-accessible isovector form factors through the decomposition
\begin{align}
    \label{eq:formfactors}
    \bra{N,\bm{p'},s'}V_\mu(0)\ket{N,\bm{p},s}
        &= \bar{u}(\bm{p'},s')\left[ \gamma_\mu F_1(Q^2) + i\frac{\sigma_{\mu\nu}q^\nu}{2m_\mathrm{N}}F_2(Q^2)\right] u(\bm{p},s),\\
    \bra{N,\bm{p'},s'}A_\mu(0)\ket{N,\bm p,s}
    &= \bar{u}(\bm{p'},s')\left[ \gamma_5\gamma_\mu G_\mathrm{A}(Q^2) + \frac{\gamma_5q_\mu}{2m_\mathrm{N}}G_\mathrm{P}(Q^2)\right] u(\bm p,s),
\end{align}
where $Q^2=-(p'-p)^2>0$ is the (spacelike) momentum transfer and $\gamma_\mu$ are the generators of the Minkowski Dirac algebra, or the commonly-used Sachs parameterization of the form factors, via
\begin{align}
    G_\mathrm{E}(Q^2) &= F_1(Q^2) - \frac{Q^2}{4m_N^2}F_2(Q^2), \\
    G_\mathrm{M}(Q^2) &= F_1(Q^2) + F_2(Q^2).
\end{align}
These quantities are also amenable to numerical calculations on a finite Euclidean lattice, and therefore provide a good test that the finite-size, discretization and excited-state effects are under control.
However, reproducing the experimental value of axial coupling of the nucleon $g_\mathrm{A}/g_\mathrm{V}=1.2723(23)$~\cite{Agashe:2014kda}, where $g_\mathrm{A}=G_\mathrm{A}(0)$ and $g_\mathrm{V}=G_\mathrm{E}(0)$, has proved challenging~\cite{Green:2014vxa} for lattice calculations.
In section~\ref{sec:methodology} we recapitulate the methodology used in this work and present results on the nucleon isovector axial coupling and electric radius in section~\ref{sec:results}.

\section{Methodology}
\label{sec:methodology}
In this work we use ensembles generated by the Coordinated Lattice Simulations (CLS) effort with $N_\mathrm{f}=2+1$ dynamical flavours of non-perturbatively O($a$)-improved Wilson fermions and the tree-level Symanzik-improved gauge action~\cite{Bruno:2014jqa}.
This allows us to investigate systematic errors due to the finite lattice spacing and \-extra\-polate the results in the pseudoscalar mass to the pion mass.
The set of ensembles used in this work and the number of measurements performed are listed in table~\ref{tab:ensembles}.
The axial vector current is non-perturbatively renormalized~\cite{Bulava:2016ktf}.
We use the conserved lattice vector current, which is not O($a$)-improved, but whose charge is correctly normalized.

\subsection{Effective matrix elements}

\begin{table}
    \centering
    \begin{tabular}{ccccrr}
        \toprule
        id      & $m_\mathrm{PS}$/MeV       & $a$/fm      &$L/a$ 
            & $N_\mathrm{meas}$     
            & $t_\mathrm{s}$/fm\\
        \midrule
        H102& 350     & 0.08 & 32
            & 7988
            & $\{1.0, 1.2, 1.4\}$\\
        H105& 280     & "   & 32
            & 11428
            &"\\
        *C101    & 220     & " & 48
            & 32416
            &"\\
        \midrule
        N200      & 280     & 0.06 & 48
            & 3200
            & $\{0.8, 0.9, 1.0, 1.15, 1.3, 1.4\}$\\
        *D200      & 200     & " & 64
        & 29088
            & $\{1.0, 1.15, 1.3, 1.4\}$\\
        \bottomrule

    \end{tabular}
    \caption{Parameters of the ensembles and measurements used in this work. The truncated-solver method was used for ensembles marked with an asterisk.}
    \label{tab:ensembles}
\end{table}

In order to access the matrix elements, we compute two- and three-point functions of nucleon interpolating operators, $\bar\Psi(x)$, and a local fermion bilinear, $J(x)$,
\begin{align}
    \label{eq:corrfunctions}
    C_{2}(t;\bm p)
    &= \frac{1}{2}\mathrm{Tr}[(1+\gamma_0)(1+i\gamma_5\gamma_3)
    \sum_{\bm x} e^{-i\bm p\cdot\bm x}\bra{0}{\Psi(\bm x, t)\bar{\Psi}(0)}\ket{0}], \\
        C_{3,J}(t,t_\mathrm{s};\bm q)
        &= \frac{1}{2}\mathrm{Tr}[(1+\gamma_0)(1+i\gamma_5\gamma_3)
        \sum_{\bm x,\bm y} e^{i\bm q\cdot\bm y}\bra{0}{\Psi(\bm x, t_\mathrm{s})J(\bm y, t)\bar{\Psi}(0)}\ket{0}],
    \label{eq:nucleon_corr}
\end{align}
where $\gamma_\mu$ are the generators of the Euclidean Dirac algebra, and use the ratio
\begin{align}
    \label{eq:ratio}
    R_J(t,t_\mathrm{s};Q^2) = \frac{C_{3,J}(t,t_\mathrm{s};\bm q)}{C_2(t_\mathrm{s};\bm 0)}\sqrt{\frac{C_2(t_\mathrm{s}-t;-\bm q)C_2(t,\bm 0)C_2(t_\mathrm{s};\bm 0)}{C_2(t_\mathrm{s}-t;\bm 0)C_2(t;-\bm q)C_2(t_\mathrm{s};-\bm q)}}
\end{align}
to cancel the overlaps of the interpolating operator with the ground state, $\bra{0}\Psi(0)\ket{N,\bm p, s}$.
This ratio was observed to have reduced variance over others considered in ref.~\cite{Alexandrou:2008rp} and is related to the effective axial coupling via $g^\mathrm{eff}_\mathrm{A}(t,t_\mathrm{s})=\mathrm{Im}\, R_{A_3}(t,t_\mathrm{s},0)$, or the effective electromagnetic form factors, $G_X^\mathrm{eff}(t,t_\mathrm{s};Q^2)$, via
\begin{align}
    \label{eq:effformfactors}
    {V_0}(t, t_\mathrm{s}; Q^2) {=}
        \sqrt{\frac{m_N + \Ep}{\Ep}} G^\mathrm{eff}_\mathrm{E}(t,t_\mathrm{s};Q^2),\quad
        \mathrm{Re}\, R_{V_i}(t, t_\mathrm{s}; Q^2) {=}
        \frac{\epsilon^{ij}q^j}{\sqrt{2\Ep(\Ep+m_N)}}G^\mathrm{eff}_\mathrm{M}(t,t_\mathrm{s};Q^2),
\end{align}
where $\epsilon^{12}=+1=-\epsilon^{21}$.
The effective form factors are suitable estimators for the form factors in the regime where $t$ and $t_\mathrm{s}-t$ are large compared with $m_\mathrm{PS}^{-1}$.
The nucleon interpolating operators are constructed from Gaussian-smeared quark fields with APE smeared links~\cite{Gusken:1989qx,Albanese:1987ds}.
The smearing parameters for each lattice spacing were chosen to minimize the effective mass of the nucleon two-point function at a relatively small time separation where the contribution from the excited states is large.

\subsection{Truncated-solver method}
The analysis of Parisi~\cite{Parisi:1983ae} or Lepage~\cite{Lepage:1989hd} suggests that the signal-to-noise ratio of the nucleon correlator at large times will decay exponentially, and worsen as the pseudoscalar mass is reduced.
This hampers the computation of the effective form factors at sufficiently large source-sink separations.
Therefore, for the ensembles with small pseudoscalar masses, marked with an asterisk in table~\ref{tab:ensembles}, we use the truncated-solver method~\cite{Collins:2007mh,Bali:2009hu,Shintani:2014vja,vonHippel:2016wid}.
An estimator for fermionic observables with reduced variance is constructed by solving the Dirac equation with many source positions to a low precision.
Although this estimator is cheap to compute, it must be corrected for its bias.
Schematically, for a two-point function $\mathcal O(x,y)$, the estimator on the r.h.s.~of
\begin{align}
    \eval{\mathcal O(x,y)} &= \Eval{\frac{1}{N_{y_0}}\sum_{\{y_0\}}{\mathcal O}_\mathrm{LP}(x+y_0,y+y_0)} +
        \underbrace{\eval{\mathcal O(x,y)} - \eval{\mathcal O_\mathrm{LP}(x,y)}}_%
                   {\textrm{bias correction}},
    \label{eq:ama}
\end{align}
should have a reduced variance, where $\mathcal{O}_\mathrm{LP}(x,y)$ is the observable computed using a low-precision solution of the Dirac equation.
As long as the contribution to the variance of the bias correction is small enough, this procedure can be more efficient than using many standard sources to improve the signal.
The variance due to the correction can be tuned by adjusting the solver residue or by estimating it with more samples per configuration.

\begin{figure}[t]
    \centering
    \includegraphics[scale=0.5]{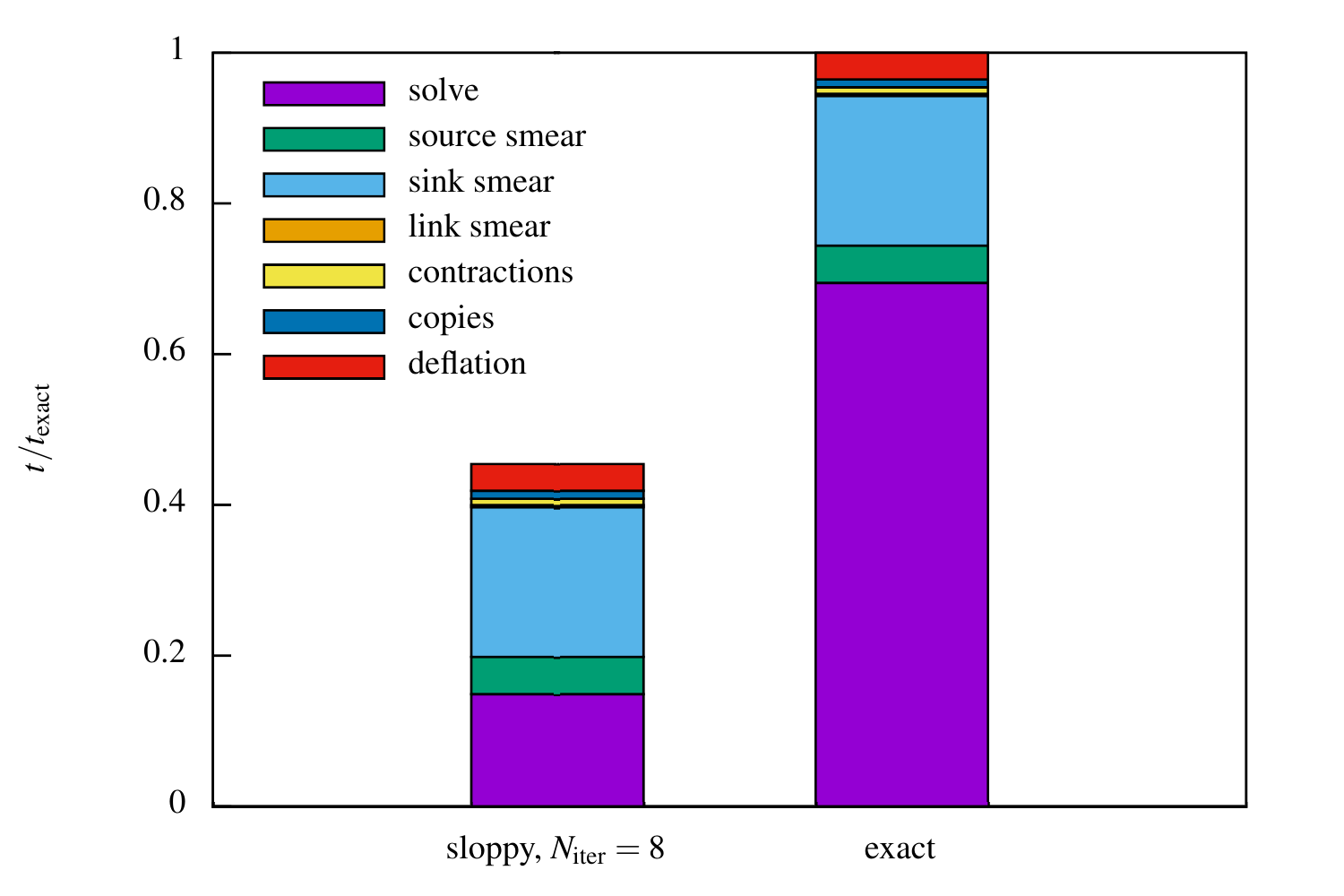}%
    \includegraphics[scale=0.5]{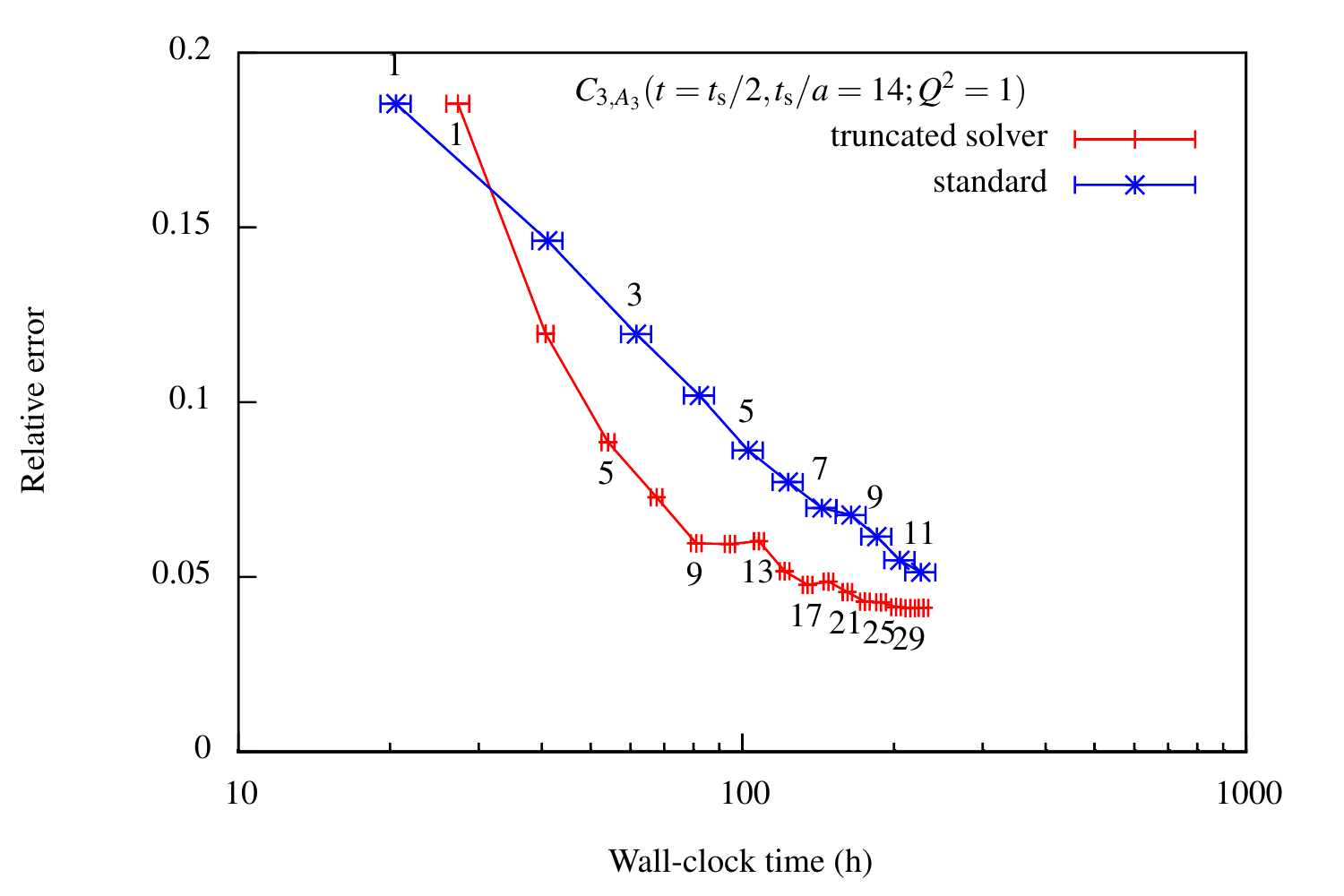}
    \caption{Left: break-down of the cost of the low- and high-precision measurements of two- and three-point functions on the C101 ensemble. Right: relative precision on the three-point function versus the cost using exact (blue) and low-precision (red) sources. The labels indicate the number of sources used per timeslice.}
    \label{fig:ama}
\end{figure}

Figure~\ref{fig:ama} (left) shows the break-down of the cost of the measurement of the two- and three-point functions for the C101 ensemble using a low-precision solution of the Dirac equation (left bar) normalized to the cost of the measurement with a ``exact'' solution of the Dirac equation (right bar).
A metric of the improvement is the cost to obtain a given statistical precision on a particular observable.
In figure~\ref{fig:ama} (right) the relative precision for the three-point function is plotted versus the cost for the truncated-solver method (red) and the standard method (blue) is shown.
Although some saturation of the relative error is observed with many source positions, the method effectively reduces the cost of the measurements.

\subsection{Extracting asymptotic matrix elements}
Although variance-reduction techniques such as the truncated-solver method can make source-sink separations in the region of 1.5~fm accessible, that may be insufficient to isolate the contribution of the ground-state matrix element with the desired accuracy.

In the case of the axial coupling, chiral perturbation theory can provide guidance on the contribution of the leading excited states to the estimator for the charge~\cite{Bar:2016uoj,Tiburzi:2015tta}.
That analysis suggests that a deviation on the order of 5\% from the asymptotic value persists at source-sink separations of $\approx1.5$~fm at the physical pion mass.
The result is independent of the choice of local interpolating operator, provided that the smearing radius is smaller than the pion correlation length.
Another scenario suggests that at accessible source-sink separations, higher two-particle states which include the effects of final-state interactions, may also contribute significantly~\cite{Hansen:2016qoz}.
These contributions cannot be estimated in chiral perturbation theory.

\begin{figure}[t]
    \centering
    \includegraphics[scale=0.7]{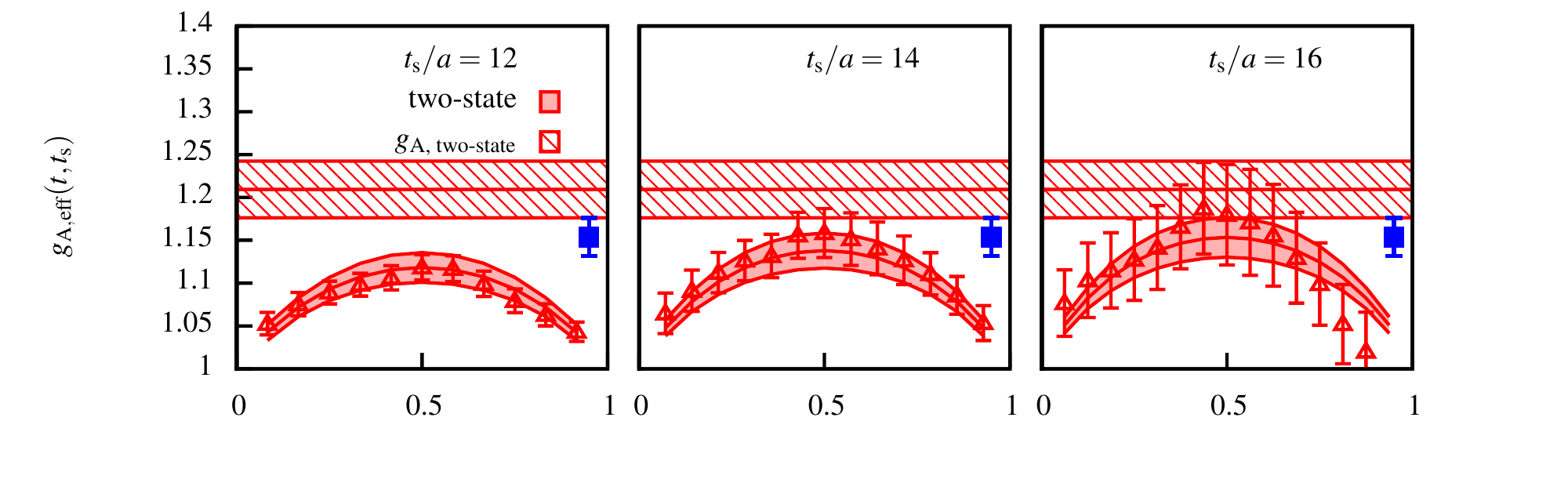}
    \includegraphics[scale=0.7]{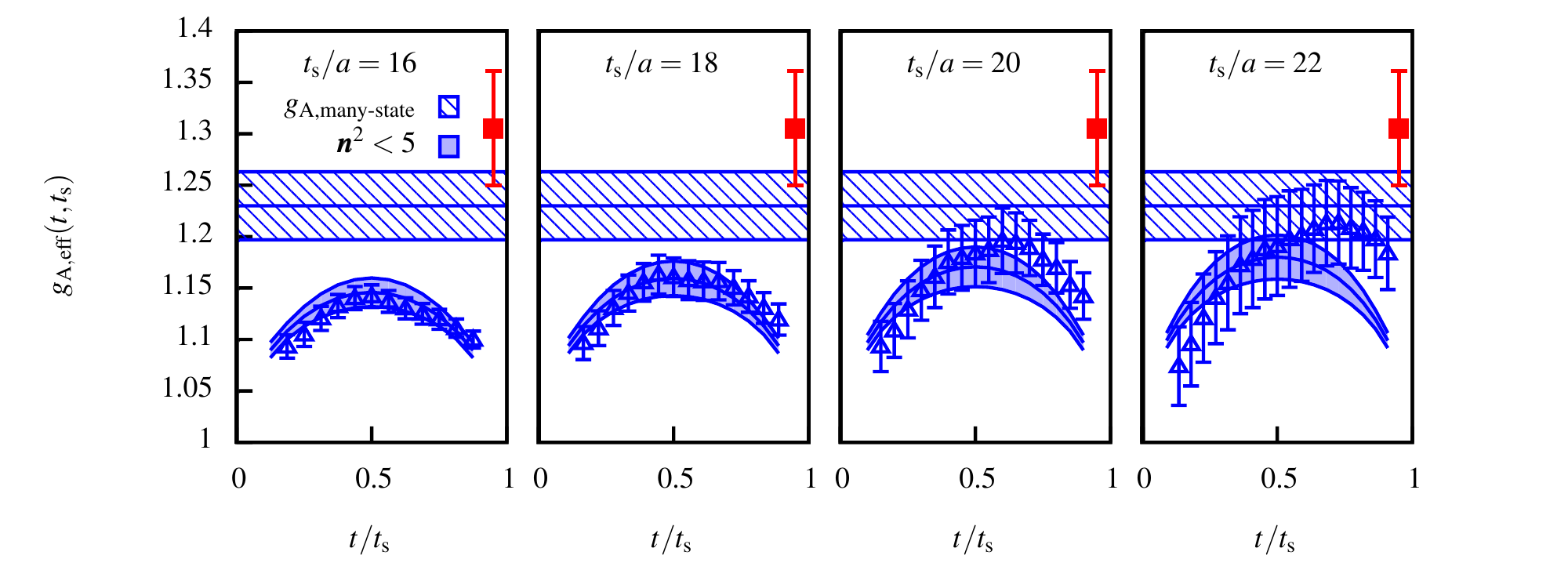}
    \caption{Top panels: effective axial matrix element for various source-sink separations and simultaneous two-state fit (shaded band) for the H105 ensemble.
    Bottom panels: analogous figure for the many-state fit on the D200 ensemble.
        The hatched band corresponds to the estimate for the asymptotic matrix element.
        For comparison, the data point on the right represents the estimate from the many-state (two-state) fit in the top (bottom) panels.
        }
    \label{fig:ga_H105}
\end{figure}
Therefore, we utilize the summation method or model the contribution of excited states explicitly in order to obtain an estimate for the ground-state matrix element in the regime where residual excited-state contamination is not negligible.
In the following, we adopt the procedures and notation of ref.~\cite{Capitani:2015sba}.
\begin{itemize}[leftmargin=*]
    \item[(i)] \emph{Summation method}: by summing over the position of the temporal coordinate of the operator insertion, an estimator with excited-state effects $\mathrm{O}(e^{-\Delta t_\mathrm{s}})$ can be obtained from the slope of the sum as a function of $t_\mathrm{s}$,
        \begin{align}
            \sum_{t=0}^{t_\mathrm{s}}G_X^{\mathrm{eff}}(t,t_\mathrm{s}; Q^2)\stackrel{t_\mathrm{s}\gg 0}{\rightarrow}
            K_X(Q^2)+t_\mathrm{s}\hat G_X(Q^2) + \ldots,
            \label{eq:summation}
        \end{align}
        where $K_X(Q^2)$ denotes a (generally divergent) constant, and elided terms are exponentially suppressed.
    \item[(ii)] \emph{Two-state fit}: the form-factor $\hat{G}_X(Q^2)$ is estimated by modelling the contribution of the leading excited state by
        \begin{align}
            G^\mathrm{eff}_X(t,t_\mathrm{s};Q^2)
            = \hat{G}_X(Q^2) +  a_X(Q^2)e^{-\Delta t} + b_X(Q^2)e^{-\Delta'(t_\mathrm{s}-t)}.
            \label{eq:two_state}
        \end{align}
        In this case, the energy gap is fixed to the lowest non-interacting level, $\Delta=m_\pi$ or $2m_\pi$ and $\Delta'=2m_\pi$ and the transition matrix elements, $a_X$, $b_X$, are left as free parameters.
        In the case of the axial coupling, the couplings are subject to the constraint $a_X=b_X$.
        Simultaneous fits are performed to all source-sink separations to extract the asymptotic value.
        In figure~\ref{fig:ga_H105} (top), the effective matrix element at three source sink separations is shown for the H105 ensemble with an uncorrelated two-state fit depicted with the band and the asymptotic value with the hatched band.
        The data show large curvature which appears to be reproduced by the model.

    \item[(iii)] \emph{Many-state fit}: for the axial coupling we also employ a variation of the two-state fit where we include only the contributions of non-interacting two-particle states with relative momentum-squared $\bm p ^2 < \bm p_\mathrm{max}^2$.
        From ref.~\cite{Hansen:2016qoz}, the non-interacting levels, $E_{\bm p}=\sqrt{\bm p^2 + m_\mathrm{PS}^2} + \sqrt{\bm p^2 + m_\mathrm{N}^2}$, are to a good approximation close to the true finite-volume two-particle levels.
        Furthermore, in LO chiral perturbation theory the accompanying matrix elements are slowly-varying with energy, which we use to constrain the corresponding fit parameters.
        The lower panel of figure~\ref{fig:ga_H105} shows the uncorrelated many-state fit including states with $\bm n^2_\mathrm{max} = \bm p_\mathrm{max}^2/(2\pi /L)^2 = 5$, which also provides a good description of the data but a systematically lower value of the asymptotic matrix element than the standard two-state fit.
\end{itemize}

\section{Conclusions}
\label{sec:results}
\begin{figure}[t]
    \centering
    \includegraphics[scale=0.5]{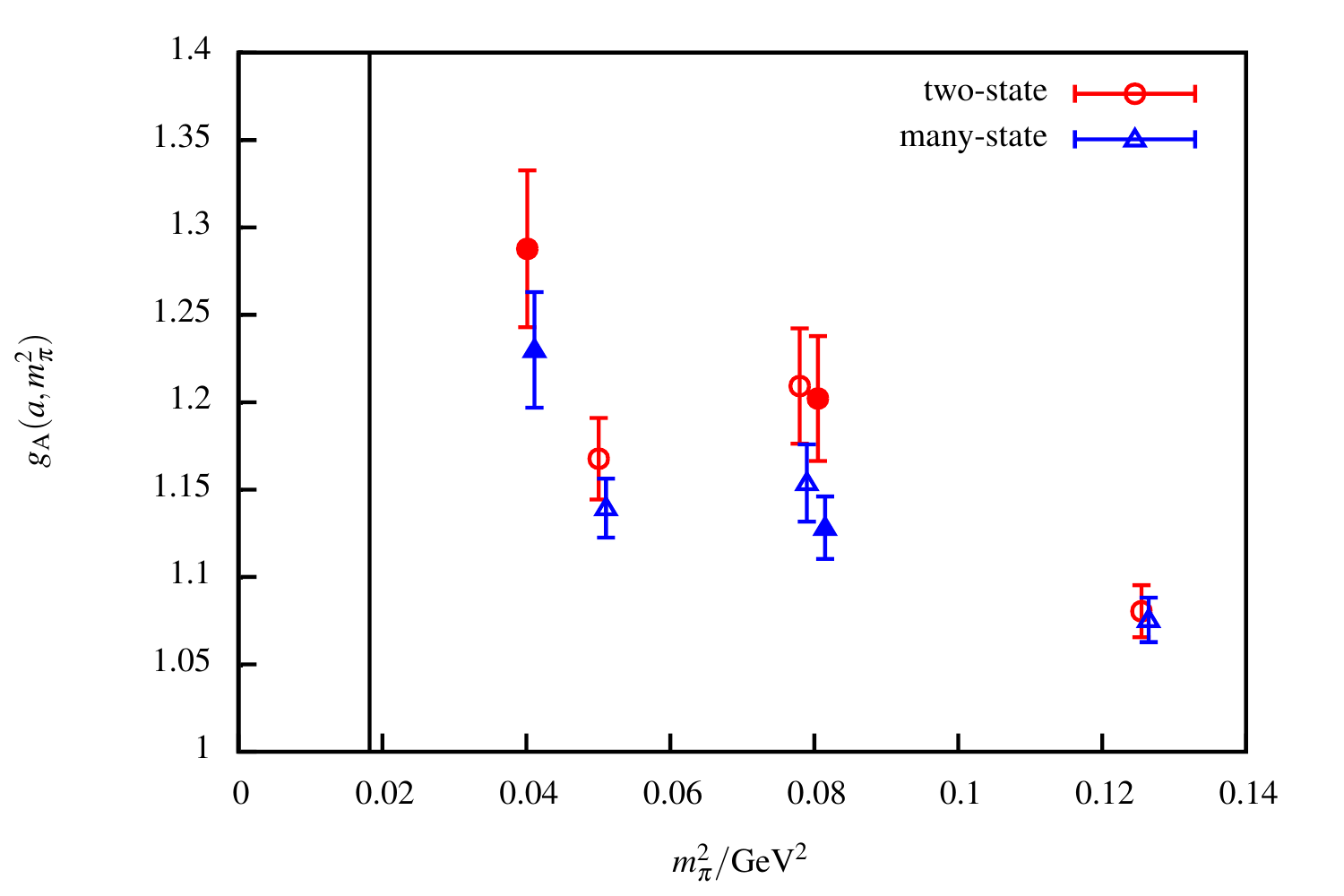}%
    \includegraphics[scale=0.5]{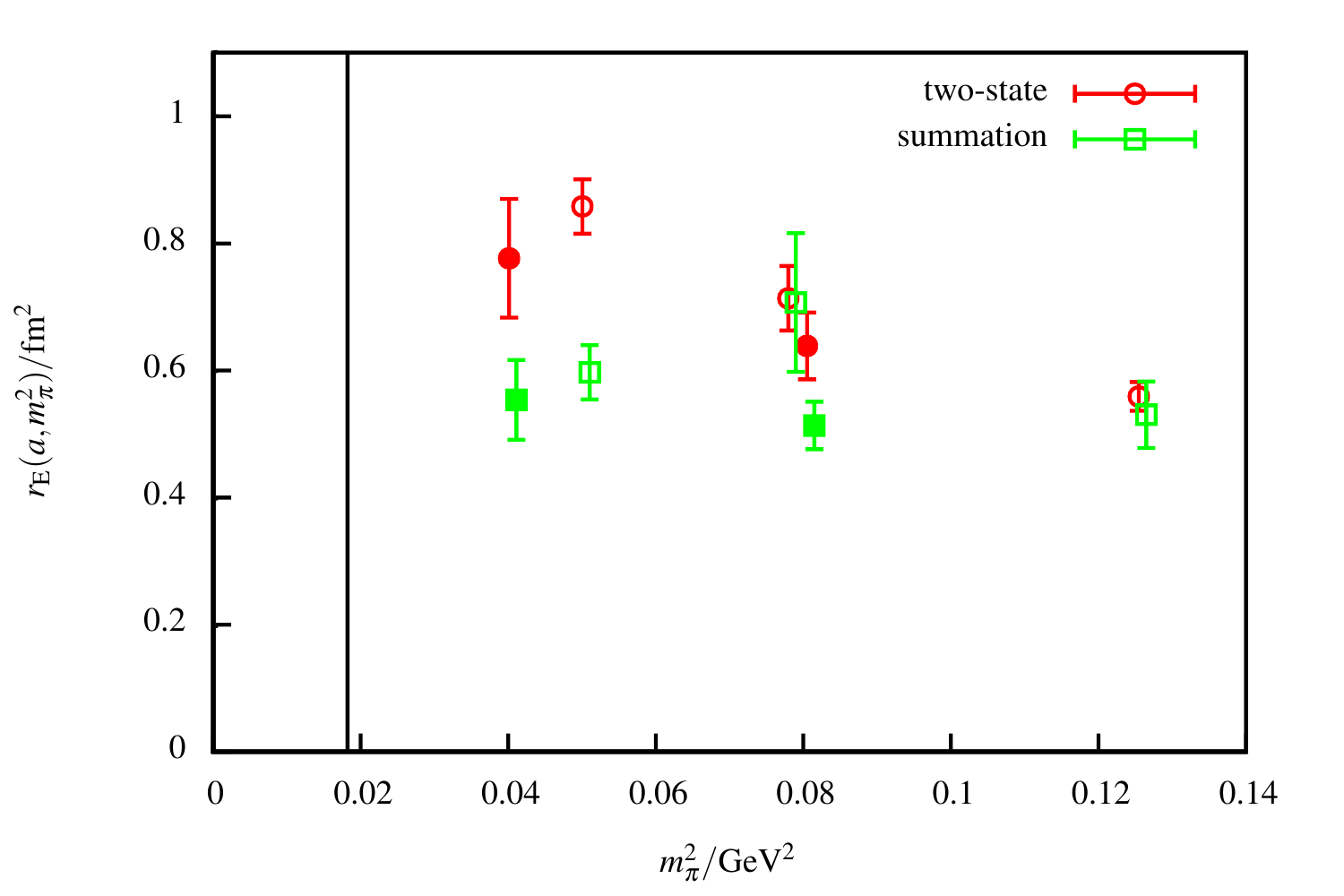}
    \caption{Left: Axial coupling of the nucleon from the two-state fit (red), many-state fit (blue) versus the pseudoscalar mass. Right: The nucleon electric radius obtained from a dipole fit to the form factor estimated using the two-state fit (red) and summation method (green). Filled and open symbols correspond to the fine and coarser lattice spacing respectively.}
    \label{fig:summary}
\end{figure}

Figure~\ref{fig:summary} (left) summarizes the estimates of the axial coupling obtained using the two-state fit and many-state fit versus the pseudoscalar mass.
The estimate from the summation method is noisier when just three source-sink separations are available and is omitted.
The estimate from the many-state fit is systematically lower than the two-state fit.

\begin{figure}[t]
    \centering
    \includegraphics[scale=0.7]{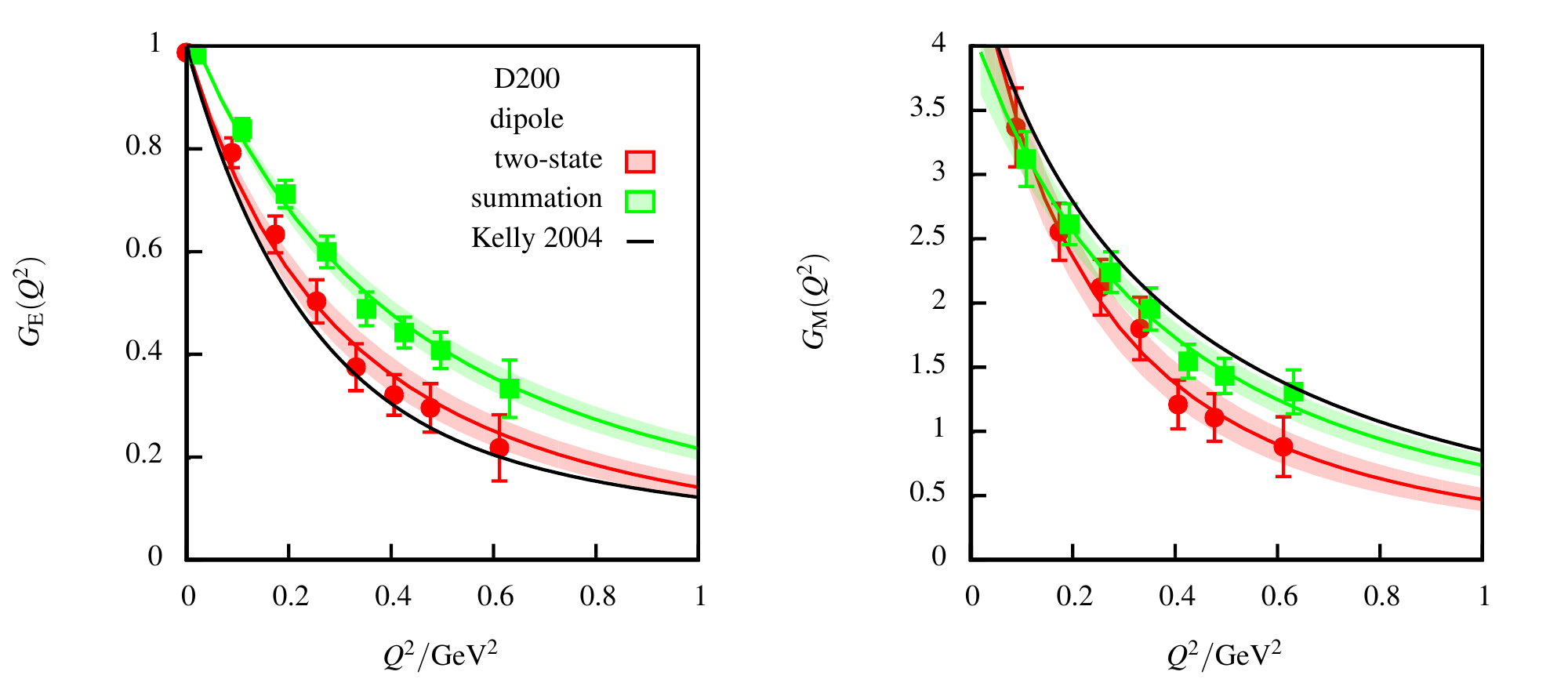}
    \caption{Electric (left) and magnetic (right) form factors for the D200 ensemble with dipole fits to the effective form factors estimated using the summation method (green) and two-state fit (red).
        The black curve is the parameterization of the experimental data from ref.~\cite{Kelly:2004hm}.}
    \label{fig:ge_D200}
\end{figure}

The electromagnetic form factors for the D200 ensemble are shown in figure~\ref{fig:ge_D200}.
At this pseudoscalar mass, a discrepancy between the methods arises when the correction from the excited state in the two-state fit becomes large, similarly to what was observed in ref.~\cite{Capitani:2015sba}.
This feature is visible in the electric radius, extracted from a dipole fit to the form factor, in the right-hand panel of figure~\ref{fig:summary}. 
Further analysis is required to better control the excited-state effects in these nucleon structure calculations.

{\small
\paragraph{Acknowledgments}
The author thanks colleagues at the University of Mainz, in particular Max Hansen and Jeremy Green, for useful discussions.
We are grateful to our colleagues within the CLS initiative for sharing ensembles.
The ensembles were partly generated at the Swiss National Supercomputing Centre, Forschungszentrum J\"ulich and the Leibniz Supercomputing Centre.
Our programmes use the \emph{QDP++} library~\cite{Edwards:2004sx} and deflated SAP+GCR solver from the \emph{openQCD} package~\cite{openQCD}.
All observables were computed on the \emph{Clover} cluster at the Helmholtz Institute Mainz.
}
\renewcommand{\normalsize}{\footnotesize}
\setlength\bibitemsep{0.5\itemsep}
\vspace{-1em}
\printbibliography{}

\end{document}